\documentclass{aa}
\usepackage{amsmath}
\usepackage{graphicx}
\usepackage{txfonts}

\begin{document}

   \title{Impactor flux and cratering on Ceres and Vesta: \\ 
          Implications for the early Solar System
   }

   \author{ G. C. de El\'\i a  
           \thanks{gdeelia@fcaglp.unlp.edu.ar}
          \and
           R. P. Di Sisto
          }

   \offprints{G. C. de El\'{\i}a
    }

  \institute{Facultad de Ciencias Astron\'omicas y Geof\'\i sicas, Universidad 
Nacional de La Plata and \\
Instituto de Astrof\'{\i}sica de La Plata, CCT La Plata-CONICET-UNLP \\
   Paseo del Bosque S/N (1900), La Plata, Argentina.
                }

   \date{Received / Accepted}


\abstract
{The forthcoming arrival of the NASA's Dawn spacecraft to Ceres and Vesta means that these unexplored 
worlds in the asteroid Main Belt are targets of several studies.
}
{We study the impactor flux and cratering on Ceres and Vesta caused by the 
collisional and dynamical evolution of the asteroid Main Belt.
}
{We develop a statistical code based on a well-tested model 
for the simultaneous evolution of the Main Belt and NEA size 
distributions. This code includes catastrophic collisions and 
noncollisional removal processes such as the Yarkovsky effect and the
orbital resonances. It is worth noting that the model assumes that the dynamical depletion of the early Main Belt
was very strong, and owing to that, most Main Belt comminution occurred when 
its dynamical structure was similar to the present one.
}
{Our results indicate that the number of $D >$ 1 km Main Belt asteroids striking Ceres and Vesta over the Solar System history are
 approximately 4\,600 and 1\,100, respectively. Moreover, the largest Main Belt asteroids expected 
to have impacted Ceres and Vesta had a diameter of 71.7 for Ceres and 21.1 km for Vesta.
As for the cratering, our simulations show that the surfaces of Ceres 
and Vesta present a wide variety of craters with different sizes. In 
fact, the number of $D >$ 0.1 km craters on Ceres is $\sim$ 
3.4 $\times$ 10$^{8}$ and  6.2 $\times$ 10$^{7}$ on Vesta. Moreover, the number of craters 
with $D >$ 100 km are 47 on Ceres and 8 on Vesta. On the other hand,
our study indicates that the $D =$ 460 km crater observed on Vesta had to be formed 
by the impact of a $D \sim$ 66.2 km projectile, which has a probability of occurr  $\sim$ 30 \% 
over the Solar System history.}
{If significant discrepancies between our results about the cratering on Ceres and Vesta and data obtained from 
the Dawn Mission were found, they should be linked to a higher degree of collisional evolution during the early Main Belt and/or 
the existence 
of the late heavy bombardment.  An increase in the collisional activity 
in the early phase may be provided for an initial configuration of the 
giant planets consistent with, for example, the Nice model. 
From this, the Dawn Mission would be able to give us clues about the initial configuration of the early Solar System 
and its subsequent dynamical evolution.}


\authorrunning{G. C. de El\'\i a \& R. P. Di Sisto
               }
\titlerunning{
Impactor flux and cratering on Ceres and Vesta
                             }

\maketitle
\section{Introduction}

Ceres and Vesta are the largest and most massive members of a vast population of small bodies located between Mars and 
Jupiter commonly called the asteroid Main Belt. Several numerical models (e.g. Bottke et al. \cite{Bottke2005a, Bottke2005b}; 
O'Brien \& Greenberg \cite{OBrien2005}; de El\'{\i}a \& Brunini \cite{deElia2007}) indicate that the size distribution of
the Main Belt asteroids is determined primarily by collisional processes. In fact, these studies suggest that most of the largest objects
($D \gtrsim$ 120 km) have never been disrupted, while many smaller asteroids are byproducts of fragmentation events among 
the largest bodies.
While Ceres and Vesta have not been targets of catastrophic collisions, they have been exposed to cratering impacts over the age of the  
Solar System. 
 In fact, the existence of the Vesta family (Binzel \& Xu \cite{Binzel1993}) is clear evidence that this object has 
undergone large cratering impacts over time.
Cratering is one of the most important processes that determine the
morphology of the surface of a Solar System object. The understanding and quantification of the impactor source
population onto an object and the observation of the object surface help for understanding the dynamical and 
physical history of both the impactor population and the target.

 Launched in September 2007,  NASA's Dawn Mission was captured in orbit by Vesta on July 15, 2011, and it should reach the 
vicinities of Ceres in February 2015.
The theoretical predictions of producing of craters may be compared with observations of Ceres and Vesta. 
This will help, on the 
one hand, for identifying the source of craters and, on the other hand, accounting for the geological processes that have acted on the 
surfaces of those bodies.
Then, it is very important to study all the possible sources of crater production on Ceres and Vesta 
 in order to estimate the total crater production 
and to contrast them with observations.

In this paper we evaluate the impactor flux and cratering on Vesta and Ceres due to the collisional and dynamical evolution of the asteroid Main Belt.
 To do this, we constructed a statistical code based on the collisional model developed by Bottke et al. 
(\cite{Bottke2005a}) with some dynamical considerations from Bottke et al. (\cite{Bottke2005b}). 
A comparison between our study and data obtained from the Dawn Mission 
may be relevant for the structure and evolution of the early Solar System.

\section{The full model}

Bottke et al. (\cite{Bottke2005a}) developed a collisional model capable of tracking the evolution of the asteroid Main Belt over the Solar
System history. At each timestep, this statistical algorithm calculates the total number of catastrophic collisions between objects residing in
 different size bins, using parameters such as the mean impact velocity $\langle V \rangle$, the intrinsic collision probability 
$\langle P_{\text{i}} \rangle$, and the impact energy required for dispersal $Q_{D}$. From this, the code computes how many objects are 
catastrophically fragmented and removed from each size bin, as well as the number of fragments resulting from those collisions, which are 
distributed in the different bins according to their sizes. 
This model assumes that all breakups occur close to the catastrophic 
disruption threshold and neglects cratering events, which
produce much less ejecta over time than catastrophic disruption events, 
and highly-energetic catastrophic disruption events, which are relatively 
uncommon.

The algorithm developed by Bottke et al. (\cite{Bottke2005a}) includes the effects of an intense period of collisional evolution in the early 
massive Main Belt. Using numerical simulations, the authors found that the net collisional activity in the Main Belt over its lifetime is the 
equivalent of $\sim$ 7.5 - 9.5 Gyr of collisional activity in the current Main Belt. This ``pseudo-time approximation'' means that the Main 
Belt population required $\sim$ 1.5 - 2 times the degree of comminution than it would have experienced if it were not initially much more massive.
It is worth noting that this model assumes that the collisional history of the Main Belt has been dominated by the same 
intrinsic collision probabilities and impact velocities as are found in the current Main Belt. Based on Petit et al. 
(\cite{Petit2001, Petit2002}), Bottke et al. (\cite{Bottke2005a}) 
 consider that the dynamical removal phase of the Main Belt was short and 
owing to that the high velocity impacts did not play an important role 
in the collisional history of the Main Belt.
From this, the collisional model from Bottke et al. (\cite{Bottke2005a}) uses the current values 
of  $\langle P_{\text{i}} \rangle$  and  $\langle V \rangle$ 
 throughout the whole simulation.

This model is able to fit a wide range of observational constraints such as the Main Belt size distribution, the number of large asteroid 
families produced by the disruption of $D >$ 100 km parent bodies over the past 3 to 4 Gyr, the existence of a $D$ = 460 km crater on the intact 
basaltic crust of Vesta, and the relatively constant crater production rate of the Earth and Moon over the last 3 Gyr.

Later, Bottke et al. (\cite{Bottke2005b}) performed a study aimed at linking the collisional history of the asteroid Main Belt to its 
dynamical excitation and depletion. This work combines the collisional evolution code created by Bottke et al. (\cite{Bottke2005a}) with 
dynamical results from Petit et al. (\cite{Petit2001}), as well as the removal of bodies from the Main Belt due to the action of resonances
and the Yarkovsky effect, which enter the near-Earth asteroid (NEA) population. This collisional and dynamical evolution code also satisfies
the above constraints and successfully reproduces the observed NEA size distribution. It is worth noting that Bottke et al. (\cite{Bottke2005b}) 
validate the pseudo-time approximation proposed by Bottke et al. (\cite{Bottke2005a}). 

\subsection{Simulation parameters}

 Here, we construct a collisional model based on the
one described in Bottke et al. (\cite{Bottke2005a}) with some dynamical
considerations taken from Bottke et al. (\cite{Bottke2005b}). From
this, we track the simultaneous evolution of both the NEA and Main Belt
populations by simulating the effects of an intense collisional
evolution in the early massive Main Belt from the pseudo-time
approximation proposed by those authors.

\begin{figure}
\centering
\resizebox{\hsize}{!}{\includegraphics{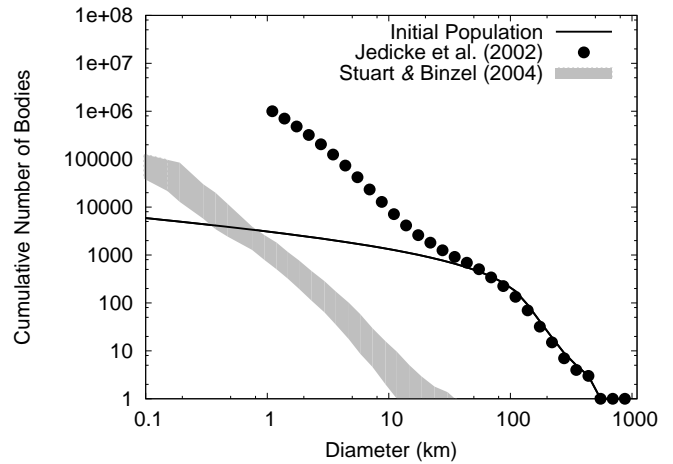}}
\caption{
 Initial Main Belt size distribution used in our simulations, the observed Main Belt size distribution 
from Bottke et al. (\cite{Bottke2005a}), which is based on Jedicke et al. (\cite{Jedicke2002}) 
with a few changes (see Sect. 2.2), and the observed NEA size distribution from Stuart \& Binzel 
(\cite{Stuart2004}).
	}
\label{Fig1}
\end{figure}

The initial population used here follows the idea proposed by Bottke et al. (\cite{Bottke2005a, Bottke2005b}). In fact, our starting size distribution
for $D>$ 200 km uses the number of observed Main Belt asteroids, with a few objects added in, to account for the Eos and Themis parent bodies.
For $\sim$ 120 $< D <$ 200 km, the population follows an incremental power-law index of -4.5, while for $D \lesssim$ 120 km an incremental power-law 
index of -1.2 is assigned. On the other hand, the NEA population always starts with no bodies. 
Figure~\ref{Fig1} shows the initial Main Belt size 
distribution used in our simulations, the observed Main Belt size distribution 
from Bottke et al. (\cite{Bottke2005a}), which is based on Jedicke et al. (\cite{Jedicke2002}) 
with a few changes, and the observed NEA size distribution from Stuart \& Binzel (\cite{Stuart2004}).
Jedicke et al. (\cite{Jedicke2002}) analyzed the absolute magnitude $H$ distribution of the Main Belt in the 
range 5.0 $< H <$ 18.5. To transform the Jedicke et al. (\cite{Jedicke2002}) $H$ distribution into a size 
distribution, Bottke et al. (\cite{Bottke2005a})  assume a visual 
geometric albedo $p_{\text{v}}$ = 0.092. Moreover, they include the 
observed asteroids for $D >$  300 km using the IRAS/color-albedo-derived
diameters cited in Farinella \& Davis (\cite{Farinella1992}).

For collisions between Main Belt asteroids, the value adopted in the collisional algorithm for the intrinsic collision 
probability is $\langle P_{\text{i}} \rangle =$ 2.86$ \times$ 10 $^{-18}$ km $^{-2}$ yr $^{-1}$  and the mean impact 
velocity is  $\langle V \rangle = $  5.3 km s$^{-1}$ (Bottke et al. \cite{Bottke1994}). To compute impacts on Ceres and 
Vesta, we use the particular values of $\langle P_{\text{i}} \rangle$  derived by O'Brien et al. 
(\cite{OBrien2011}) for each of these bodies. The values adopted for $\langle P_{\text{i}} 
\rangle$  are 3.70 $\times$ 10$^{-18}$  km$^{-2}$ yr$^{-1}$ for Ceres  and 2.97 $\times$ 
10$^{-18}$  km $^{-2}$ yr $^{-1}$ for Vesta. As for the disruption law, we use the impact energy required for 
dispersal $Q_{D}$ derived by Benz \& Asphaug (\cite{Benz1999}) for basalt at 3 km\,\,s $^{-1}$. While 
the mean impact velocity  $\langle V \rangle$ for collisions between Main Belt asteroids is 5.3 km s$^{-1}$, we adopt the 
 $Q_{D}$ law for 3 km s$^{-1}$ since Bottke et al. (\cite{Bottke2005a}) show that it produces very good fits to the 
observational constraints. 

The differential fragment size distribution (FSD) 
produced by each catastrophic disruption event is represented by
\begin{equation}
dN = BD^{-p} dD,
\label{FSD}   
\end{equation}
 where $D$  is the diameter,  $\text{d}N$  is the number of fragments in
the size range ($D$,$D+\text{d}D$), $B$  is a constant, and  $p$  is the
power-law index. The FSDs used here are based on those observed in
asteroid families like Themis or Flora (Bottke et al. \cite{Bottke2005a}).
According to Tanga et al. (\cite{Tanga1999}), the Themis family was produced by the 
super-catastrophic breakup of a  $D =$  369 km body, while the Flora family was created 
by the barely-catastrophic breakup of a  $D =$  164 km body. Bottke et al. (\cite{Bottke2005a}) show that FSDs
resulting from super-catastrophic disruption events are 
represented well by a single  $p$  value, although this is not true for FSDs 
produced by barely-catastrophic breakups. From this, Bottke et al. (\cite{Bottke2005a})
developed two different FSDs to describe the possible outcomes of a
catastrophic collision.
On the one hand, for  $D >$  150 km disruption events, the diameter of 
the largest remnant is assumed to be 50 \% the diameter of the parent body.
Moreover, the power-law index  $p$  of the differential FSD between the 
largest remnant and 1/60 the diameter of the parent body is -3.5, while 
the $p$  value is -1.5 for smaller fragments. On the other hand, for disruption events
among $D <$  150 km bodies, the largest remnant is 80 \% the diameter of 
the parent body. Moreover, the power-law index $p$  of the differential FSD between the 
largest remnant and 1/3 the diameter of the parent body is -2.3. In this case, the FSD
has a second break at 1/40 the diameter of the parent body. The  $p$
value between 1/3 and 1/40 the diameter of the parent body is -4, while
 $p$ is assumed to be -2 for smaller fragments.

On the other hand, since the NEA population is several orders of magnitude smaller than the Main Belt population, we neglect 
collisions between NEAs and between NEAs and Main Belt asteroids. In fact, we consider that the NEA population is sustained by the input of material
from the Main Belt via orbital resonances and the Yarkovsky effect. Here, we use the non-collisional removal rate of asteroids from the Main Belt 
proposed by Bottke et al. (\cite{Bottke2005b}).
Thus, bodies removed from the Main Belt by noncollisional 
processes are placed in the NEA population, which is assumed to decay with 
a mean dynamical lifetime of 5 Myr.

\begin{table}
\begin{minipage}[t]{\columnwidth}
\caption{ Main Belt model parameters (Bottke et al. \cite{Bottke2005a}).}
\label{Tabla1}
\centering
\renewcommand{\footnoterule}{} 
\begin{tabular}{c|c|c|c}
\hline \hline 
$H$ & $D$ & $dN$ & $dN_{\text{Fam}}$ \\
\hline \hline
3.25 & 980.9 & 1.0 & - \\
3.75 & 779.2 & 0.0 & - \\
4.25 & 618.9 & 0.0 & - \\
4.75 & 491.6 & 2.0 & - \\
5.25 & 390.5 & 1.0 & 1 \\
5.75 & 310.2 & 3.0 & 1 \\
6.25 & 246.4 & 8.0 & 1 \\
6.75 & 195.7 & 17.0 & 5 \\
7.25 & 155.5 & 38.0 & 5 \\
7.75 & 123.5 & 64.0 & 5 \\
8.25 & 98.1 & 91.0 & -\\
8.75 & 77.9 & 116.0 & -\\
9.25 & 61.9 & 164.0 & -\\
9.75 & 49.2 & 185.0 & -\\
10.25 & 39.1 & 224.0 & -\\
10.75 & 31.0 & 338.0 & -\\
11.25 & 24.6 & 554.0 & -\\
11.75 & 19.6 & 789.7 & -\\
12.25 & 15.6 & 1548.0 & -\\
12.75 & 12.4 & 2992.3 & -\\
13.25 & 9.81 & 5671.8 & -\\
13.75 & 7.79 & 10463.9 & -\\
14.25 & 6.19 & 18630.7 & -\\
14.75 & 4.92 & 31739.6 & -\\
15.25 & 3.91 & 51398.5 & -\\
15.75 & 3.10 & 78939.8 & -\\
16.25 & 2.46 & 115400.3 & -\\
16.75 & 1.96 & 162026.4 & -\\
17.25 & 1.55 & 221080.1 & -\\
17.75 & 1.23 & 296503.1 & -\\
18.25 & 0.98 & 394278.9 & -\\
\hline \hline

\end{tabular}
\end{minipage}
\vspace{0.2cm}

 $H$  is the absolute magnitude, $D$  the central diameter (in kilometers) 
of the bin for a visual geometric albedo  $p_{\text{v}} =$  0.092, 
 $dN$  the incremental number of asteroids in each bin based on Jedicke et al. 
(\cite{Jedicke2002}) with a few modifications (see Sect. 2.2), and 
 $dN_{\text{Fam}}$  the number of observed asteroid families in each 
bin.

\end{table}
\subsection{Runs}

We perform our numerical simulations using a pseudo-time of 9.5 Gyr. In addition, 
 since the range of 7.5 to 9.5 Gyr is what is found by Bottke et al. 
(\cite{Bottke2005a}) to give the best fits, we decide 
to carry out numerical simulations for a pseudo-time of 7.5 Gyr in order to test the sensitivity of our results to this parameter. This point is referred in Sect. 3.

On the other hand, to simulate the disruption events in the asteroid Main Belt in a more realistic manner, our algorithm works stochastically treating 
the breakups as Poisson random events (Press et al. \cite{Press1989}). A stochastic code produces different results using different seeds for the random
number generator. Thus, we develop a large number of runs using different random seeds and then interpret the results statistically.

To obtain a quantitative measure of how good a run reproduces observational data, we follow the procedure described by Bottke et al. (\cite{Bottke2005a}).
First, the metric used to determine the goodness of fit between the observed Main Belt size distribution ($N_{\text{MB}}$) and the model results 
($N_{\text{MODEL}}$) is given by
\begin{equation}
\psi_{\text{SFD}}^{2} = \sum_{D} \left(\frac{N_{\text{MODEL}}(D) - N_{\text{MB}}(D)}{0.2N_{\text{MB}}(D)} \right)^{2},
\end{equation}
where the summation extends over 1 $< D <$ 1\,000 km size bins. Second, a standard $\chi^{2}_{\text{FAM}}$ test is used to compare the number of breakups 
produced in each $D >$ 100 km size bin over the last 3.5 Gyr to the observed number of asteroid families in the 100 $< D <$ 400 km size bins.
Table~\ref{Tabla1} summarizes the observed Main Belt model parameters necessary to develop these  tests (Bottke et al. \cite{Bottke2005a}).
Following Bottke et al. (\cite{Bottke2005b}), we assume that a run produces a positive match if 
$\psi_{\text{SFD}}^{2} < 20$ and $\chi^{2}_{\text{FAM}} < 2\sigma$ (i.e., probability $>$ 5 $\%$).

Given the stochastic nature of our code, we carried out 10\,000 numerical simulations using a pseudo-time of 9.5 Gyr, 533 of which show positive matches. 
From such simulations, we calculated the flux of Main Belt asteroids of different sizes onto Ceres and Vesta. Besides using a suitable expression 
we also computed the crater diameters produced by Main Belt asteroids on such bodies. Below, we discuss the algorithms 
that relate crater diameter to impactor diameter. 

\subsection{Cratering laws on Ceres and Vesta}

Cratering processes have been extensively study through impact and explosion experiments
(e.g. Schmidt \& Housen \cite{SH87}). In a recent work interpreting the 
observations of the Deep Impact event, Holsapple \& Housen (\cite{HH07}) have updated the 
 crater scaling laws for different materials. The impact cratering scaling laws 
depend in general on two exponents $\mu$ and $\nu$ and a constant $K_1$
that characterize the different materials. The diameter 
$D_s$ of a crater produced by an impactor of diameter $d$ can be obtained from 
Holsapple \& Housen (\cite{HH07}) from the equation: 
\begin{equation}
D_s = K_1 \left[ \left(\frac{g d}{2 v_i^2}\right) 
\left(\frac{\rho_t}{\rho_i}\right)
^{2 \nu / \mu} +  \left(\frac{Y}{\rho_t v_i^2}\right)^{(2+\mu)/2} 
\left(\frac{\rho_t}{\rho_i}\right)^{\nu (2+\mu)/ \mu} \right 
]^{-\mu/(2+\mu)}   d,
\label{ds}
\end{equation}  
where $\rho_t$ is the target density, $g$ its superficial gravity,  
$Y$ its strength, $\rho_i$ the density of the impactor, and $v_i$ the impactor 
velocity.  

To calculate the craters on Ceres and Vesta, we have to take 
 their composition and superficial characteristics into account. 
Table~\ref{Tabla2} shows the values of several physical parameters 
for Ceres and Vesta. We have no 
direct information on the surfaces of those asteroids. For Ceres, observations from their 
reflectance spectra plus theoretical analysis are consistent with a rocky surface (Zolotov \cite{Z09}).
The low density of Ceres implies that it contains low-density 
compounds. The internal structure and the amounts of water in Ceres 
have already been modeled and studied (McCord \& Sotin \cite{MS05}, Zolotov \cite{Z09}), but there 
is no unique, generally accepted model for Ceres interior. Those 
models may really be tested with the observations of the Dawn Mission in 2015. 
Vesta has a basaltic surface and is considered a differentiated body (McCord 
et al. \cite{Mc70}). Then, Ceres and Vesta can be assimilated to the cratering  law that 
corresponds to wet soils and rock. For this material $\mu=0.55$, $\nu=0.4$, and $K_1 = 
0.93$. Then, replacing these values in Eq. (\ref{ds}) we can obtain the   
final diameter of the crater for a given impactor diameter.

\begin{table}
\begin{minipage}[t]{\columnwidth}
\caption{
Values of the mass $M$, the density $\rho$ (Baer \& Chesley 
\cite{Baer2008}), the diameter $D$ (Thomas et al. \cite{Thomas1997b}, \cite{Thomas2005}), 
and the surface gravity $g$ used for Ceres and Vesta. }
\label{Tabla2}
\centering
\renewcommand{\footnoterule}{} 
\begin{tabular}{l|c|c}
\hline \hline 
 & Ceres   &  Vesta  \\
\hline
$M$ (g) &  9.45 $\times$ 10$^{23}$   &  2.67 $\times$ 10$^{23}$   \\ 
$\rho$ (g cm$^{-3}$) &  2.09   &  3.42     \\
$D$ (km) & 952.4 & 516 \\
$g$ (cm s$^{-2}$) & 27.81  &  26.73   \\
\hline
\end{tabular}
\end{minipage}
\end{table}

Equation (\ref{ds}) gives a general law for impacts, and it is a 
convenient empirical smoothing function to span the transition between the gravity regime 
and the strength regime (Holsapple \cite{Holsapple1993}). 
We use this general form, since in the low gravity of asteroids 
the strength regime can be important for the smaller craters. 
The strength of the target $Y$ is a value that have many measures for a 
geological material. Housen \& Holsapple (\cite{HH03}) show that the crater size law 
depends on the average of the tensile and compressive strength. Then, we adopt a 
value of the strength $Y$ corresponding to soft rocks of 3 MPa as in 
Holsapple \& Housen (\cite{HH07}).

 Obtained from Eq. (\ref{ds}), $D_s$ is the diameter of a simple crater, 
but for a large crater, gravitational forces modify the crater form leading to complex 
craters. The transition from simple to complex craters is an observable quantity unique for each world. 
Schenk et al. (\cite{Schenk2004}) summarize results for the single-to-complex 
transition diameters for silicate planets and icy satellites. From 
this, an inverse correlation between transition diameter and surface 
gravity is evident, with icy satellites offset toward lower transition 
diameters. Since the surface gravities of Ceres and Vesta are comparable (Table~\ref{Tabla2}), 
extrapolation of the values associated with the 
silicate planets leads to a transition diameter $D_{t}$ for those objects of $\sim$ 50 km.
When we use the diameter of the complex crater from McKinnon et al. (\cite{McKinnon2003}), 
 then the final crater diameter is given by
\begin{xalignat}{4}
D &= D_s                       &&\text{for} && D_s < D_t,  \nonumber \\
D &=  1.17 D_s (D_s / D_t)^{0.13}    &&\text{for} && D_s > D_t.
\label{dfinal}
\end{xalignat}

\begin{figure}
\centering
\resizebox{\hsize}{!}{\includegraphics{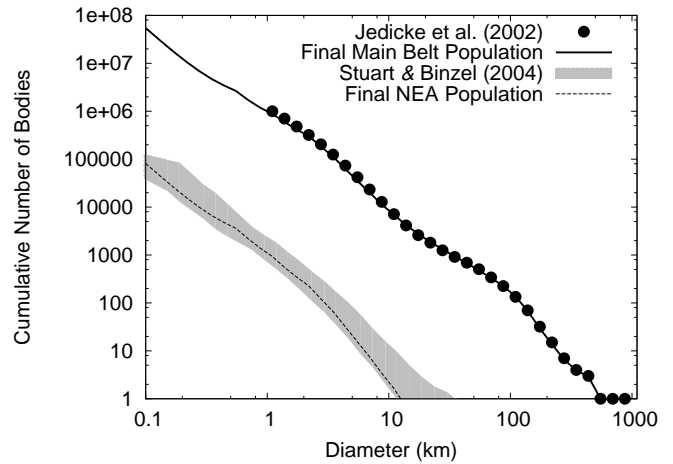}}
\caption{
Our estimates of the Main Belt (solid line) and NEA (dashed line) size distributions 
for a pseudo-time of 9.5 Gyr.
Moreover, the observed Main Belt size distribution from Bottke et al. (\cite{Bottke2005a}), 
which is based on Jedicke et al. (\cite{Jedicke2002}) with a few changes,
and the observed NEA size distribution from Stuart \& Binzel (\cite{Stuart2004}) 
are also shown.}
	
\label{Fig3}
\end{figure}

\section{Results}

We present here our main results for the impactor flux and cratering onto Ceres and Vesta 
caused by the collisional and dynamical evolution of the asteroid Main Belt.

The statistical analysis developed in this section is based on the 533 runs that produce positive matches with the observational data. In 
fact, Figure~\ref{Fig3} shows our estimates of the Main Belt and NEA cumulative size distributions for a pseudo-time of 9.5 Gyr, obtained 
from the 533 runs mentioned above. 
The results obtained from the simulations using a pseudo-time of 7.5 Gyr
do not show any significant changes in the Main Belt and NEA size 
distributions and the formation of the large asteroid families. As for the impactor flux on Ceres and Vesta, the 
number of impacts on each of these bodies estimated using a pseudo-time of 7.5 Gyr is 70 to 80 \%
 what is obtained with a 
pseudo-time of 9.5 Gyr.

\subsection{Impactor flux onto Ceres and Vesta}

\begin{figure}
\centering
\resizebox{\hsize}{!}{\includegraphics{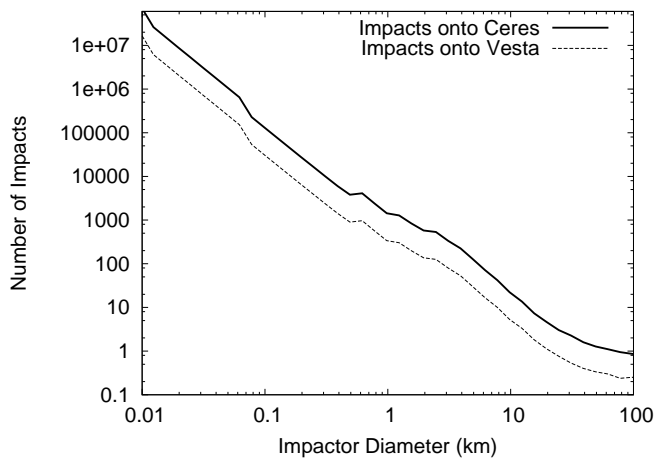}}
\caption{
Incremental number of Main Belt impactors onto Ceres (solid line) and Vesta 
(dashed line) as a function of impactor diameter obtained using a pseudo-time 
of 9.5 Gyr.
}
\label{Fig4}
\end{figure}

Figure~\ref{Fig4} shows the total number of Main Belt asteroid impacts onto Ceres 
and Vesta as a function of impactor diameter obtained using a pseudo-time of 9.5 Gyr.
These curves are computed from a simple average of the 533 runs that produce positive 
matches with the observational data.
Over the age of the Solar System, our results indicate that the number of $D >$ 1 km 
Main Belt asteroids striking Ceres is 4\,631 and 1\,096 for Vesta. 
Moreover, the largest Main Belt asteroids expected to have 
impacted Ceres and Vesta over the history of the Solar System have had  diameters of 
71.7 and 21.1 km.

\subsection{Cratering on Ceres and Vesta}

Using these impactor fluxes, it is possible to obtain the number of craters on Ceres and Vesta as a function of crater diameter.
To do this, we make use of Eqs. (\ref{ds}) and (\ref{dfinal}) with the values of the parameters $K_{1}$, $\nu$, $\mu$, $Y$, and $D_{t}$ specified in 
Section 2.4, and the physical data shown in Table~\ref{Tabla2}. 
Moreover, a density $\rho_{\text{i}} =$ 2.7 g cm$^{-3}$ is assumed for the 
impactors, which is consistent with the densities of several S-type 
asteroids (e.g. Belton et al. \cite{Belton1995}, Veverka et al. \cite{Veverka2000}).
On the other hand, the impactor velocity $v_{\text{i}}$ is given by
\begin{equation}
v_{\text{i}} = (U^{2} + v_{\text{esc}}^{2})^{1/2},
\label{v_imp}
\end{equation}
where $U$ is the hyperbolic encounter velocity, which is equal to the 
mean impact velocity $\langle V \rangle$,
and $v_{\text{esc}}$ the escape velocity of the target.
 O'Brien et al. (\cite{OBrien2011}) derived particular values of $\langle V \rangle$ 
 equal to 4.79 km s $^{-1}$ for Ceres and 4.74 km s$^{-1}$ for Vesta.
According to the physical data shown in Table~\ref{Tabla2}, Ceres and
Vesta have escape velocities of 0.51 km s$^{-1}$ and 0.37 km s$^{-1}$.
Using these values in Eq. (\ref{v_imp}), the impactor velocities
used for Ceres is 4.82 km s$^{-1}$ and 4.75 km s$^{-1}$ for Vesta.

\begin{figure}
\centering
\resizebox{\hsize}{!}{\includegraphics{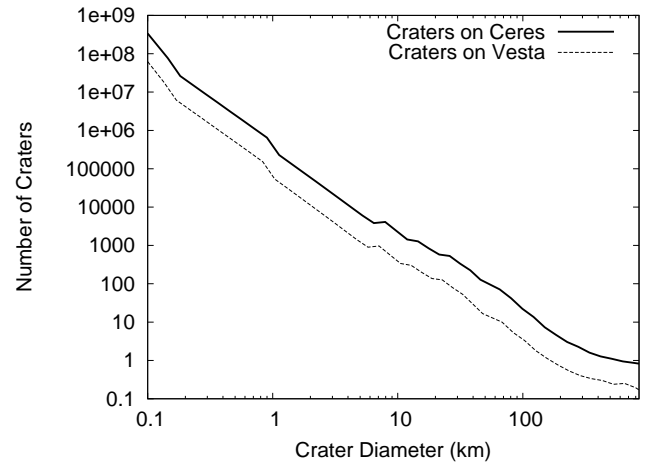}}
\caption{Incremental number of craters on Ceres (solid line) and Vesta 
(dashed line) produced by Main Belt asteroids as a function of crater 
diameter, obtained using a pseudo-time of 9.5 Gyr.}
\label{Fig5}
\end{figure}

Figure~\ref{Fig5} shows the total number of craters on Ceres and Vesta produced by Main Belt asteroids as a function of crater diameter
obtained using a pseudo-time of 9.5 Gyr. Moreover, Table~\ref{Tabla3} presents a quantitative summary of our results concerning the 
craters formed on Ceres and Vesta. From this, the surfaces of these objects should show impact structures with a wide range of sizes. 
On the one hand, our results indicate that the number of $D >$ 0.1 km craters on Ceres and Vesta 
are $\sim$ 3.4 $\times$ 10$^{8}$ and 6.2 $\times$ 10$^{7}$. On the other hand, 
our simulations show that 47 craters with $D >$ 100 km are present on Ceres and 8 on Vesta.

Hubble Space Telescope (HST) observations of Vesta have revealed a singular crater
with a diameter of about 460 km near the south pole (Thomas et al. \cite{Thomas1997a}). 
Using a two-dimensional hydrocode, Asphaug (\cite{Asphaug1997}) studied the impact origin of Vesta 
family and concluded that the 460 km crater was formed by an impactor of $\sim$ 42 km in diameter at 5.4 km s$^{-1}$.
Recently, Ivanov et al. (\cite{Ivanov2011}) have analyzed the formation of the south pole impact crater 
on Vesta from projectiles with an impact velocity of 5.5 km s$^{-1}$ 
and sizes ranging from 40 to 96 km in diameter. Their results suggest the 
possibility that the projectile diameter may be larger than $\sim$ 40 km.

Using Eqs. (\ref{ds})  from Holsapple \& Housen (\cite{HH07}) and (\ref{dfinal})
from McKinnon et al. (\cite{McKinnon2003}), the transition 
diameter specified for Vesta in Section 2.4, and the physical data shown in Table~\ref{Tabla2}, we conclude that a 
$D$ = 42 km impactor produces a transient crater of $D \sim$ 217 km and a final complex crater of $D \sim$ 307 km on Vesta\footnote{
It is worth noting that, since the $D =$  460 km singular crater 
is comparable to the Vesta diameter, the cratering laws 
used here may break down in this regime.}. 
From this, the transient crater diameter is $\sim$ 0.7 of the final crater diameter.  
Several works such as those by Grieve et al. (\cite{Grieve1981}), Melosh 
(\cite{Melosh1982}), and Croft (\cite{Croft1985}), suggest that the transient crater diameter 
is estimated at 0.5 to 0.65 of the final crater diameter. However, these 
 studies are based on lunar and terrestrial impact structures so 
we think that it would not be appropriate to compare them with our analysis 
relative to Vesta.

Taking  Eq. (\ref{dfinal}) from McKinnon et al. (\cite{McKinnon2003}) into account 
for complex craters, a final crater with a diameter of $\sim$ 460 km is formed by a $D =$ 42 km 
projectile if the transition diameter $D_{t}$ is of 2.2 km, which is a 
typical value for Venus and Mars. However, from Schenk et al. 
(\cite{Schenk2004}), we consider this value of the transition diameter to be too small for Vesta.
Either way, the Dawn Mission data will be crucial for estimating the transition 
diameter for asteroid-sized objects.

Using the most appropriate value of 50 km for the transition diameter $D_{t}$, 
our study indicates that the 460 km crater observed on Vesta's surface had to be formed 
by a projectile of $\sim$ 66.2 km in diameter, which is consistent with 
suggesting by Ivanov et al. (\cite{Ivanov2011}). Our results indicate that 
such an event has a probability of occurring of $\sim$ 30 \% over the Solar System history.
We discuss this in the next section.
 
\begin{table}
\begin{minipage}[t]{\columnwidth}
\caption{
Craters on Ceres and Vesta.
Total number of $D >$ 0.1, 1, 10, and 100 km craters on Ceres and Vesta 
produced by Main Belt asteroids using a pseudo-time of 9.5 Gyr.
}
\label{Tabla3}
\centering
\renewcommand{\footnoterule}{} 
\begin{tabular}{l|c|c}
\hline \hline 
 & Ceres   &  Vesta  \\
\hline 
$N(D > 0.1$ km)  &  3.37 $\times$ 10$^{8}$   &  6.24 $\times$ 10$^{7}$ \\
$N(D > 1$ km)    & 576\,178  &   112\,653  \\
$N(D > 10$ km)   &  6\,291   &  1\,216  \\
$N(D > 100$ km)  & 47  &  8    \\
\hline
\end{tabular}
\end{minipage}
\end{table}

\section{Discussion}

Bottke et al. (\cite{Bottke2005a}) created a numerical model capable of 
tracking how the Main Belt population was affected by collisional 
evolution from the end of accretion among $D <$ 1000 km objects to the 
present time. These authors find that the net collisional activity in the Main Belt over its lifetime is the 
equivalent of $\sim$ 7.5 - 9.5 Gyr of collisional activity in the current Main Belt.
In fact, the extra comminution had to come from a collisional phase 
occurring early in  Solar System history when there were many more $D <$ 1000 km 
objects in the primordial Main Belt than in the current 
population. Finally, the extra material in this population would have 
been primarily removed by dynamical processes rather than collisional 
evolution.  

It is worth noting that there are some caveats and limitations to the model that should be mentioned. 
Bottke et al. (\cite{Bottke2005a}) assume that most Main Belt comminution occurred when the intrinsic collision
probability $\langle P_{\text{i}} \rangle$ and the mean impact velocity $\langle V \rangle$ were comparable to their current 
values. This result is based on the dynamical simulations of the early Main Belt from Petit et al. (\cite{Petit2001}).
These authors developed a dynamical evolution model of the Main Belt, 
with Jupiter and Saturn on their current orbits. With this orbital configuration for the gas giant planets, 
Petit et al. (\cite{Petit2001}) show that the depletion of asteroid 
Main Belt was very strong, removing $\sim$ 90 \% of the original population 
in less than 10 Myr after the formation of Jupiter. Based on this model, Bottke et al. (\cite{Bottke2005a}) assume that the dynamical removal phase of the Main Belt was short 
enough that its evolution has been dominated by the same collisional 
parameters as found in the current Main Belt. Later, Bottke et al. 
(\cite{Bottke2005b}) combined dynamical results from Petit et al. (\cite{Petit2001}) 
and the collisional code of Bottke et al. (\cite{Bottke2005a}) to model 
the evolution of the Main Belt over the age of the Solar System. Unlike 
Bottke et al. (\cite{Bottke2005a}), the algorithm developed by Bottke et al. 
(\cite{Bottke2005b}) incorporates time-varying collisional parameters 
such as the intrinsic collision probability $\langle P_{\text{i}} 
\rangle$  and the mean impact velocity  $\langle V 
\rangle$. From this model, Bottke et al. (\cite{Bottke2005b}) 
find that the number of breakups occurring over 4.5 Gyr of evolution is 
consistent with the results of Bottke et al. (\cite{Bottke2005a}) using 
the pseudo-time approximation. Moreover, they find that the high velocity
impacts do not appear to have a dominant effect on Main Belt comminution,
mainly because these velocities become important
only after the excited population has been significantly depleted.

However, the depletion rate of the early asteroid Main Belt depends 
on the initial orbital configuration of the giant planets. The Nice 
model from Tsiganis et al. (\cite{Tsiganis2005}) 
strongly suggests that the initial system of outer planets had a more compact configuration 
(all within $\sim$ 15 AU of the Sun), with nearly circular and coplanar orbits. 
In the context of this model, O'Brien et al. (\cite{OBrien2007}) 
studied the primordial excitation and clearing of the asteroid Main 
Belt, assuming Jupiter and Saturn on initially circular orbits. A 
relevant result is that the excitation and depletion of the asteroid 
Main Belt derived from the Nice model are slower than those found by Petit et al. 
(\cite{Petit2001}) with Jupiter and Saturn on their current orbits. In 
fact, 90 \% of the original population is removed from the Main Belt in $\sim$ 70 Myr. 
Since in this scenario the massive population of the primitive Main Belt is removed more slowly, 
the degree of collisional evolution may have been greater than  derived from the model of 
Bottke et al. (\cite{Bottke2005a}). Therefore, the degree 
of collisional evolution of Ceres and Vesta could provide clues to the initial orbital 
configuration of the gas giant planets.

On the other hand, the model from Bottke et al. (\cite{Bottke2005a}) does not include the collisional effects of external 
impactors not belonging to the Main Belt. The so-called late heavy bombardment (LHB) is a period of timeh
during which it is believed that a large number of objects from the outer Solar System impacted the terrestrial 
planets and the Main Belt asteroids about 3.9 Gyr ago. This event may have led to significant effects on the 
degree of collisional evolution suffered by the Main Belt. It is worth noting that the existence of the LHB is
still being debated in the literature (Chapman et al. \cite{Chapman2007}).

In particular, we estimate that Vesta had a $\sim$ 30 \%  probability of being impacted by a $D =$ 66.2 km projectile, which is 
capable of producing the singular crater of 460 km in diameter observed on its surface. 
 A slower asteroid removal rate slower that is consistent with the Nice model 
and/or the existence of the LHB could 
lead to a higher degree of collisional evolution for the Main Belt increasing then the probability 
of producing the crater of 460 km on Vesta. However, a  $\sim$ 30 \% probability of 
occurrence does not allow us to discard the production 
of this crater in the present collisional evolution model. If the  $D =$ 
 460 km crater on Vesta formed during the early massive phase of 
the Main Belt, this would be consistent with studies developed by 
Bogard \& Garrison (\cite{Bogard2003}), who propose that Vesta suffered such huge 
impact 4.48 Gyr ago.

According to the comments made in this section, we believe that, if significant discrepancies between our results concerning 
the collisional history of Ceres and Vesta and the observational data obtained from the Dawn Mission were found, they should be linked to
a higher  degree of collisional evolution in the Main Belt. An increase in the collisional activity may be due to
\begin{itemize}
\item[1-]  a different initial configuration of the giant planets 
consistent with, for example, the Nice model, and/or
\item[2-] the existence of the LHB.
\end{itemize}
If such discrepancies are eventually substantiated, future works should focus on accurately quantifying the degree of collisional evolution suffered 
by the early massive Main Belt as well as during the LHB, assuming that the outer planets had initial orbits similar to and substantially 
different from (e.g. Nice model) their current ones.
From this, the Dawn Mission would be able to give us interesting clues about the initial 
configuration of the early Solar System and its subsequent dynamical evolution.

\section{Conclusions}

We have presented a study aimed at analyzing the impactor flux and cratering on Ceres and Vesta produced by Main Belt asteroids.
To do this, we constructed a statistical code based on Bottke et al. (\cite{Bottke2005a, Bottke2005b}) that includes catastrophic collisions and 
noncollisional removal processes, such as the Yarkovsky effect and the orbital resonances. Assuming that the dynamical depletion of 
the early Main Belt was very strong, and owing to that, most Main Belt comminution occurred when the intrinsic collision probability 
$\langle P_{\text{i}} \rangle$ and the mean impact velocity $\langle V \rangle$ were comparable 
to their current values. Our main results are the following
\begin{itemize}
\item The number of $D >$ 1 km Main Belt asteroids striking Ceres and Vesta over the Solar System history is approximately 4\,600 and 1\,100.
\item The largest Main Belt asteroids expected to have impacted Ceres and Vesta had diameters of 71.7 and 21.1 km.
\item The surfaces of Ceres and Vesta  should show a wide variety of craters with a wide range of sizes. On the one hand, the number of $D >$ 0.1 km craters on Ceres and Vesta are $\sim$ 3.4 $\times$ 10$^{8}$ and  6.2 $\times$ 10$^{7}$. On the other hand, 
the surfaces of Ceres and Vesta present 47 and 8 craters with $D >$ 100 km.
\item Using an appropriate value of 50 km for the transition diameter from simple to complex craters together with the expressions derived by
Holsapple \& Housen (\cite{HH07}) and McKinnon et al. (\cite{McKinnon2003}), 
the $D =$ 460 km crater observed on Vesta had to be formed by a $D \sim$ 66.2 km projectile.
\item Such an event has an occurrence  probability $\sim$ 30 \% over the Solar System history. 
 Then, we cannot discard the production of this crater in the present collisional evolution 
model. An asteroid removal rate that is slower is consistent with the Nice model and/or the
existence of the LHB which could lead to a higher degree of collisional evolution 
of the Main Belt thereby increasing the probability of producing the crater of 460 km on Vesta.
\item We suggest that, although the impact that formed the  $D =$ 
460 km crater on Vesta cannot be discarded in the present collisional evolution model, it is more 
likely to have occurred in the 
early massive phase of Main Belt while it was being excited and depleted or during
the LHB. This conclusion may become consistent with studies from Bogard \& Garrison (\cite{Bogard2003}) who propose that such an impact occurred 
4.48 Gyr ago.
\item We believe that if significant discrepancies between our results about the cratering on Ceres and Vesta and data obtained from 
the Dawn Mission were found, they should be linked to a higher degree of collisional evolution during the early massive phase of the Main 
Belt and/or the existence of the LHB. From this, the Dawn Mission could play an important role by providing information on the 
formation and evolution of the early Solar System.

\end{itemize}

\begin{acknowledgements}
We acknowledge  Francisco Azpilicueta for valuable discussions during this work. 
 We also thank referee David P. O'Brien for his helpful and constructive 
reviews.
\end{acknowledgements}

\end{document}